# Electronics Instrumentation for the Greenland Telescope

Derek Kubo[1a], Chih-Chiang Han[a], Hiroaki Nishioka[a], Ryan Chilson[a], Ranjani Srinivasan[a], Sheng-Feng Yen[a], Kuo-Chieh Fu[a], Homin Jiang[a], Kuan-Yu Liu[a], Ta-Shun Wei[a], Chih-Wei Huang[a], Chen-Yu Yu[a], Peter Oshiro[a], Shu-Hao Chang[a], Chung-Cheng Chen[a], Philippe Raffin[a], Yau-De Huang[a], Pierre Martin-Cocher[a], Ming-Tang Chen[a], Makoto Inoue[a], Satoki Matsushita[a], Keiichi Asada[a], Shoko Koyama[a], Patrick Koch[a], Paul T. P. Ho[a], Yang-Tai Shaw[a], Timothy J. Norton[b], Nimesh A. Patel[b], Shepherd S. Doeleman[b], Daniel Bintley[c], Craig Walther[c], Per Friberg[c], Jessica Dempsey[c], Hideo Ogawa[d], Kimihiro Kimura[d], Yutaka Hasegawa[d], Ching-Tang Liu[e], Kou-Chang Han[e], Song-Chu Chang[e], Li-Ming Lu[e]

[a]Academia Sinica, Institute of Astronomy & Astrophysics, No 1, Sec 4, Roosevelt Rd, Taipei 10617, Taiwan, ROC; [b]Harvard-Smithsonian Center for Astrophysics, 60 Garden St, Cambridge, MA 02138, USA; [c]East Asian Observatory, 660 N Aohoku Pl, Hilo, HI 96720, USA; [d]Osaka Prefecture University, 1-1 Gakuen-cho, Naka-ku, Sakai, Osaka, 599-8531, Japan; [e]National Chung-Shan Institute of Science and Technology, No 300-5, Ln 277, Xian St, Xitun Dist, Taichung City, 40722, Taiwan, ROC.

**ABSTRACT**

The Greenland Telescope project has recently participated in an experiment to image the supermassive black hole shadow at the center of M87 using Very Long Baseline Interferometry technique in April of 2018. The antenna consists of the 12-m ALMA North American prototype antenna that was modified to support two auxiliary side containers and to withstand an extremely cold environment. The telescope is currently at Thule Air Base in Greenland with the long-term goal to move the telescope over the Greenland ice sheet to Summit Station. The GLT currently has a single cryostat which houses three dual polarization receivers that cover 84-96 GHz, 213-243 GHz and 271-377 GHz bands. A hydrogen maser frequency source in conjunction with high frequency synthesizers are used to generate the local oscillator references for the receivers. The intermediate frequency outputs of each receiver cover 4-8 GHz and are heterodyned to baseband for digitization within a set of ROACH-2 units then formatted for recording onto Mark-6 data recorders. A separate set of ROACH-2 units operating in parallel provides the function of auto-correlation for real-time spectral analysis. Due to the stringent instrumental stability requirements for interferometry a diagnostic test system was incorporated into the design. Tying all of the above equipment together is the fiber optic system designed to operate in a low temperature environment and scalable to accommodate a larger distance between the control module and telescope for Summit Station. A report on the progress of the above electronics instrumentation system will be provided.

## 1. INTRODUCTION

The Greenland Telescope (GLT) is a collaborative project of Academia Sinica Institute of Astronomy & Astrophysics (ASIAA) of Taiwan and the Smithsonian Astrophysical Observatory (SAO) of the USA. The GLT antenna consists of the 12-m ALMA North American prototype antenna manufactured by Vertex Antennentechnik GmbH and was awarded to SAO in 2011. Subsequent to the procurement of the antenna, ASIAA with support from SAO coordinated the design modifications and retrofitting to support two auxiliary side containers and to withstand the extremely cold environment

---

[1] For further information contact Derek Kubo, email dkubo@asiaa.sinica.edu.tw, telephone 1-808-961-2929

of Summit Station [1]. The development of the electronics instrumentation system took place concurrently with the antenna modifications and is partitioned into the following areas: receiver; local oscillators and references; intermediate frequency processing and down conversion; digital backend; calibration and diagnostics; and fiber optics. This electronics instrumentation system was completed during the fall of 2017 and temporarily installed into the 15-m JCMT (James Clerk Maxwell Telescope) in Hawaii to perform design verification tests prior to proceeding to Greenland. Interferometric fringes were successfully obtained between the JCMT and SAO's Submillimeter Array during September of 2017. After completion of this milestone, the electronics instrumentation system was shipped to Thule and installed into the GLT antenna and associated trailers. A report on the progress and performance of the above electronics instrumentation system is provided herein.

## 2. RECEIVERS

The GLT receiver consists of a suite of specialized heterodyne instruments that in conjunction with the primary mirror determine the overall sensitivity of the telescope and is the subject of a separate report [2]. A brief description of the GLT receiver system is presented here to provide context to the detailed electronic instrumentation that follows. Figure 1 describes receiver that consists of a single cryostat that houses three separate cartridges to support observation bands of 86 GHz, 230 GHz, and 345 GHz. The receiver-1 cold cartridge assembly (CCA) was designed by Osaka Prefecture University and supports circular polarization and sideband separation with four 4-8 GHz intermediate frequency (IF) outputs. The associated warm cartridge assembly (WCA) is a modified ALMA band-6 unit and provides the final local oscillator to the mixers as well as IF amplification. The receiver-2 CCA and WCA are the ALMA band-7 design that support linear polarization with four 4-8 GHz IF outputs. This receiver-2 will utilize an external ¼ wave-plate (not shown in figure) to convert the two linear polarizations to circular. And finally the receiver-3 CCA is based on a W-band circularly polarized receiver that was developed by ASIAA for the Yuan-Tseh Lee Array telescope. The associated WCA for this is a modified ALMA band-6 unit with two 4-8 GHz IF outputs.

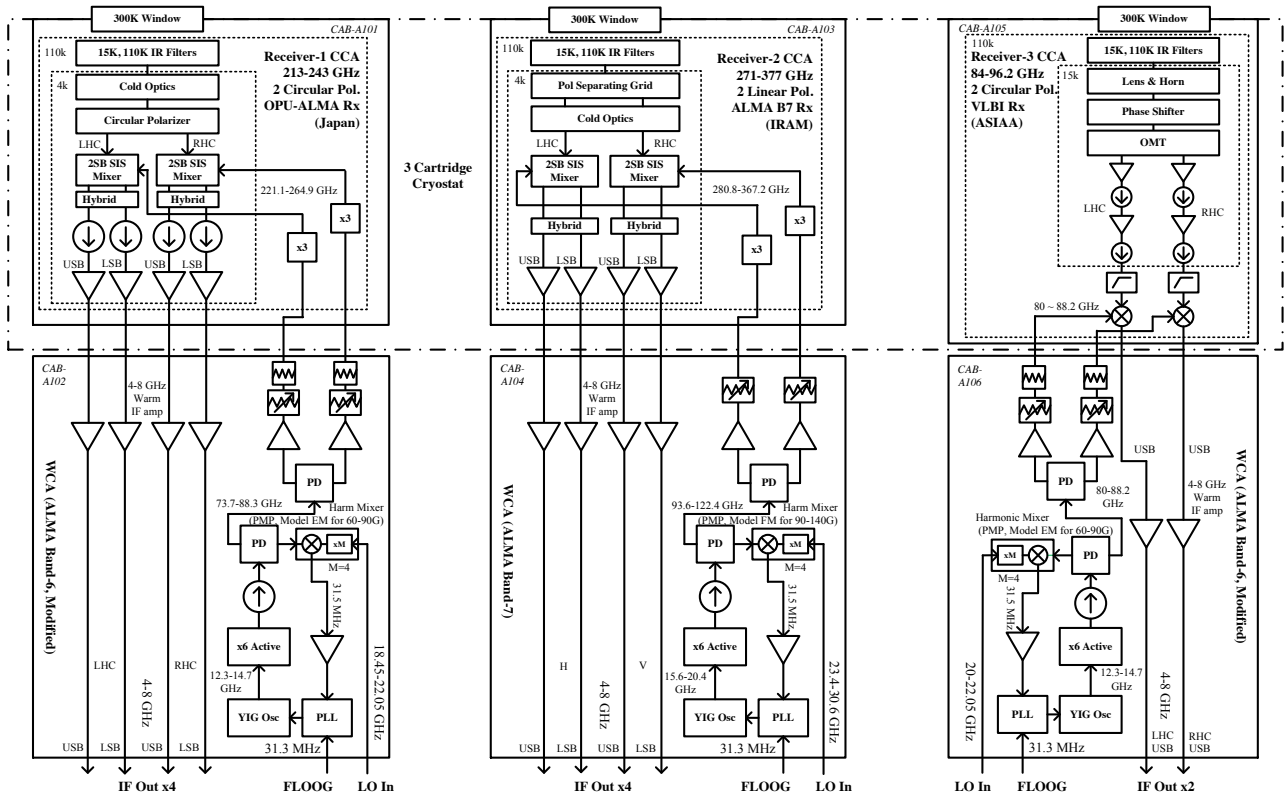

Figure 1. Simplified functional block diagram of receiver system. Three separate CCAs support frequency bands that cover 213-243 GHz, 271-377 GHz, and 84-96.2 GHz. The WCAs utilized the ALMA band-6 and band-7 designs.

# 3. LOCAL OSCILLATORS AND REFERENCES

The fundamental source for the local oscillators (LOs) generated within the GLT system is the T4 Science iMaser 3000 instrument package that produces 5 MHz, 10 MHz, 100 MHz, and 1-PPS outputs [3]. The maser is housed within a modified shipping container that provides active temperature control and is situated in an area away from vibration and magnetic fields. Figure 2 represents the functional block diagram of the LO subsystem. A Keysight E8257D RF synthesizer with low phase noise option UNX serves as the primary high frequency LO source for the receivers [4]. This LO is combined with the maser 100 MHz reference and modulated for optical fiber transmission using an Optilab LT-40 transmitter. Though the 10 MHz output from the maser has better phase noise performance than the 100 MHz output, the 100 MHz was used for optical transmission to avoid the undesired low frequency roll-off effects presented by the optical hardware. An erbium doped fiber amplifier (EDFA) is used to overcome optical component and connector losses prior to the Optilab LR-30 optical receiver located in the antenna receiver cabin. The two dense wave division multiplexer (DWDM) devices are used to separate out wavelengths for round-trip phase monitoring described later in this paper. The output of the Optilab LR-30 optical receiver is filtered to separate the 100 MHz signal from the 18-31 GHz LO signal. The 100 MHz signal from the maser is phase locked to a 10 MHz crystal oscillator and serves as the master 10 MHz reference within the antenna receiver cabin for secondary LO and clock generation.

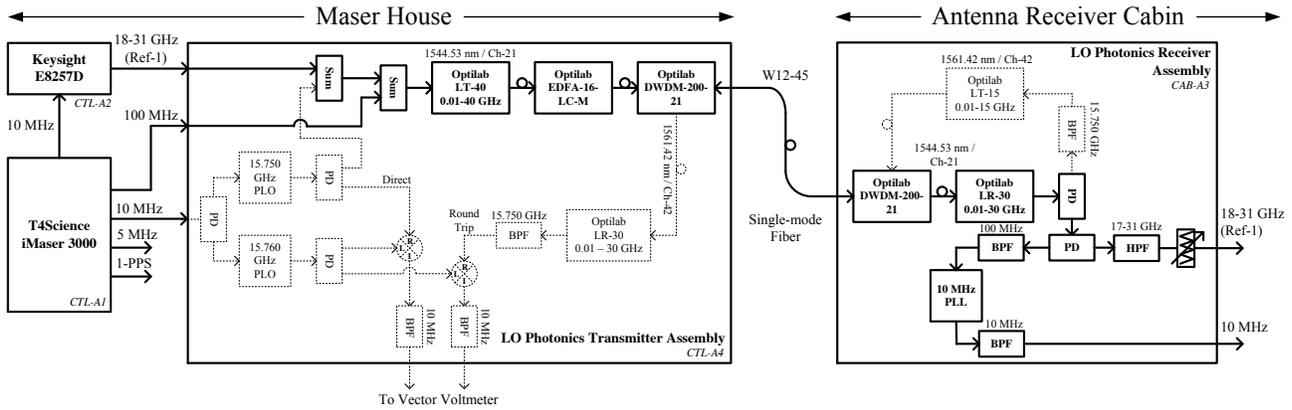

Figure 2. Local oscillator generation and transmission. A low phase noise RF synthesizer locked to the maser 10 MHz reference is used to generate a variable frequency 18-31 GHz LO required by the receiver WCAs. This RF signal is combined with the 100 MHz maser signal and optically modulated for transmission over optical fiber to the telescope receiver cabin. The optical signal is demodulated in the receiver cabin and separated out into its original constituent signals. A 10 MHz crystal oscillator is phase locked to the maser 100 MHz to serve as the master 10 MHz reference within the antenna receiver cabin.

A prototype LO photonics transmitter assembly shown in Figure 3 was fielded in Thule Air Base during November of 2017 and continues to operate as of this writing. The interior of the maser house is shown in Figure 4 with the maser instrument encased in a servo controlled thermal enclosure provided by T4 Science. The Keysight E8257D synthesizer and photonics transmitter assembly are installed in the equipment rack. A completed LO photonics receiver assembly shown in Figures 5 and 6 was installed into the antenna receiver cabin during the same time period.

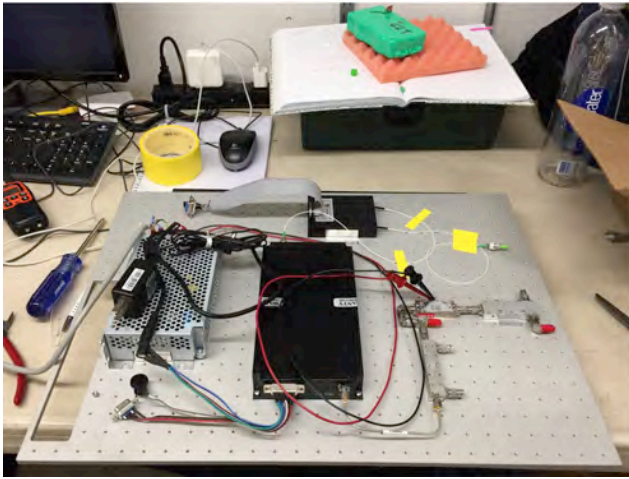

Figure 3. Prototype LO photonics transmitter assembly. This unit does not include the round trip phase monitoring hardware.

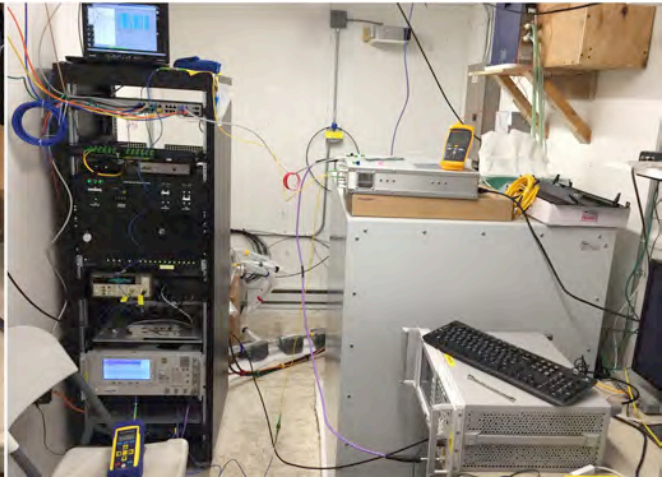

Figure 4. Maser house with equipment rack on left and T4 Science maser contained within thermal enclosure on right.

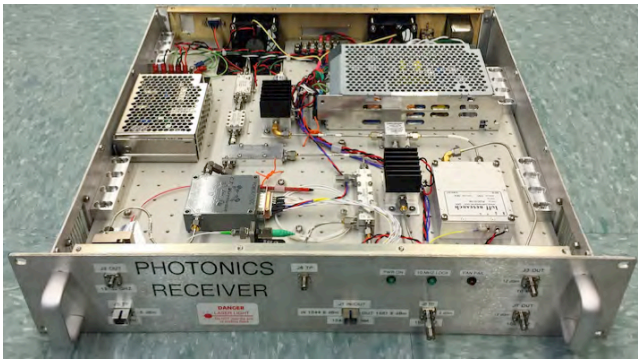

Figure 5. LO photonics receiver assembly. Top view of DC power supplies and RF components, includes round trip phase monitor hardware.

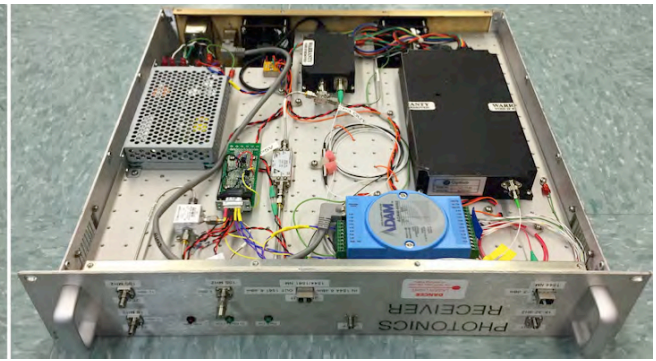

Figure 6. LO photonics receiver assembly. Bottom view of DC power supply, RF components, optical receiver and transmitter, and monitor and control hardware.

### 3.1 Performance of 10 MHz Reference in Receiver Cabin

The spectral performance of the reference signals were characterized before and after optical transmission using a Keysight N9010A spectrum analyzer. Figures 7 and 8 represent the measured 100 MHz signals from the maser and photonics receiver assembly, respectively, using a span of 1 MHz, resolution bandwidth (RBW) of 1 KHz, and video bandwidth (VBW) of 100 Hz. Note the addition of the undesired 150 kHz spurs after transmission over the fiber optic link and was isolated to the LT-40 transmitter. Figure 9 and 10 represent the 10 MHz signals from the maser and photonics receiver assembly, respectively, using the same span and RBW/VBW settings. Note that the output shown in Figure 10 is after the 100 MHz to 10 MHz phase locked loop (PLL) and effectively removes the undesired 150 kHz spurs. Finally, Figures 11 and 12 represent the close-in look at the same 10 MHz signals using a span of 100 Hz and RBW /VBW settings of 1 Hz. Under these settings it becomes quite clear how good the maser performance is with a signal-to-noise pedestal ratio of 90 dB. The reconstructed 10 MHz reference in the receiver cabin shows a degraded signal-to-noise pedestal ratio of 62 dB but has been deemed sufficiently good based on empirical results of the secondary LOs and clocks generated from this reference.

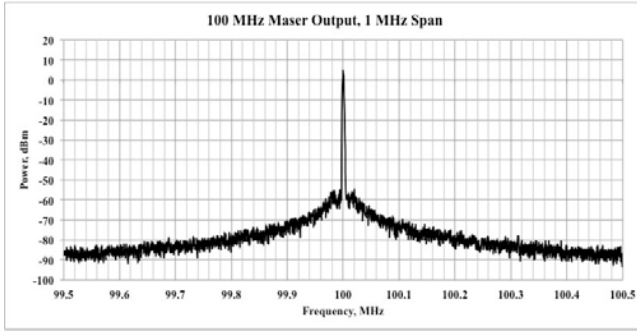

Figure 7. 100 MHz output from maser, 1 MHz span. $P_{out}$ = +4.7 dBm

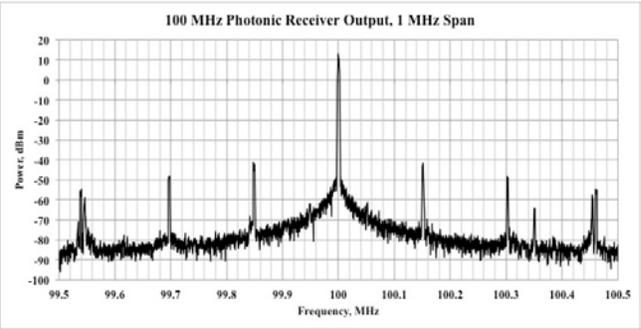

Figure 8. 100 MHz output from LO photonics receiver, 1 MHz span. Note 150 kHz spurs caused by optical transmitter. $P_{out}$ = +13.2 dBm.

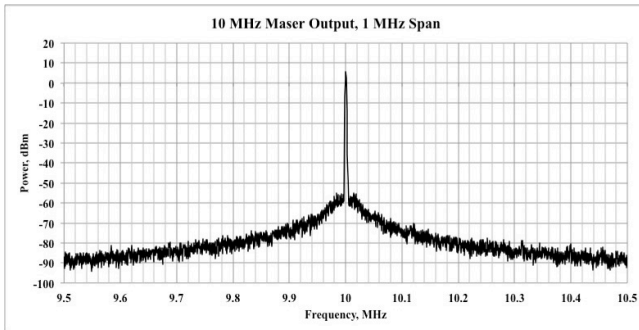

Figure 9. 10 MHz output from maser, 1 MHz span. $P_{out}$ = +5.6 dBm.

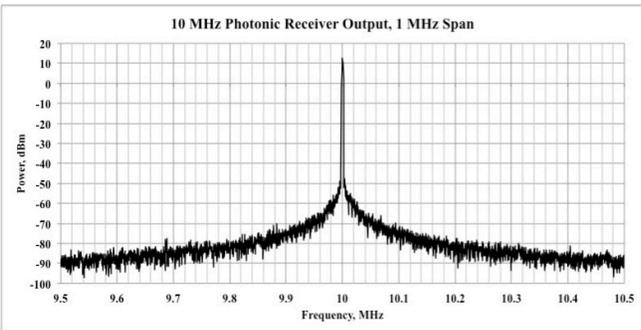

Figure 10. 10 MHz output from LO photonics receiver, 1 MHz span. Note that the 10 MHz PLL removes 150 kHz spurs. $P_{out}$ = +12.7 dBm.

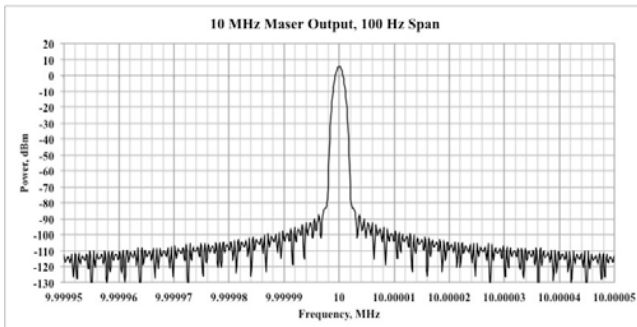

Figure 11. 10 MHz output from maser, 100 Hz span. Note that the narrow span and RBW reveals the exceptionally good spectral purity of the maser. $P_{out}$ = +5.6 dBm.

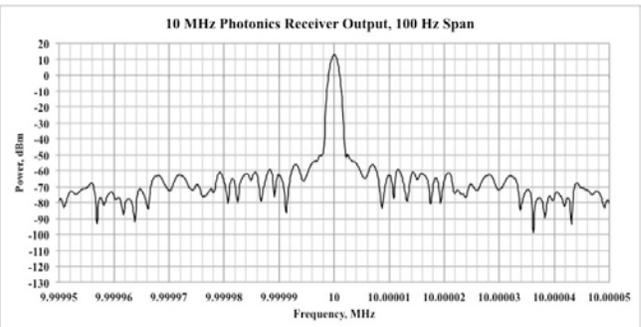

Figure 12. 10 MHz output from LO photonics receiver, 100 Hz span. Note the degraded SNR as compared to the maser output, though it is still very high at 62 dB. $P_{out}$ = +12.7 dBm.

### 3.2 Performance of 18-31 GHz LO Reference in Receiver Cabin

Similar to the previous tests, the spectral performance of the high frequency LO reference signals were characterized before and after optical transmission using the N9010A spectrum analyzer. Figures 13 and 14 represent the measured LO signals at a frequency of 26.0 GHz from the E8257D synthesizer and photonics receiver assembly outputs, respectively, using a span of 1 MHz, RBW of 1 KHz, and VBW of 100 Hz. Note the 150 kHz spurs are still present at the photonics receiver output. Figures 15 and 16 represent the close-in look at the same signals using a span of 100 Hz and RBW and VBW settings of 1 Hz. The spectral performance of the photonics receiver output is quite similar to the synthesizer output indicating very little degradation of signal quality after optical transmission.

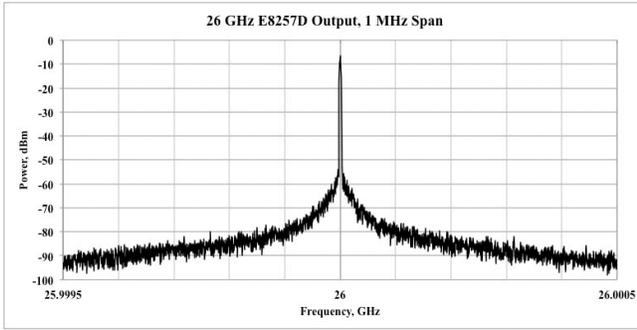
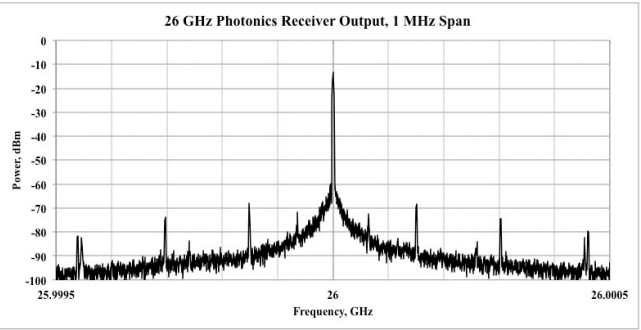

Figure 13. 26.0 GHz output from E8257D synthesizer, 1 MHz span. $P_{out}$ = -6.5 dBm.

Figure 14. 26.0 GHz output from photonics receiver assembly, 1 MHz span. Note the presence of the 150 kHz spurs. $P_{out}$ = -13.4 dBm.

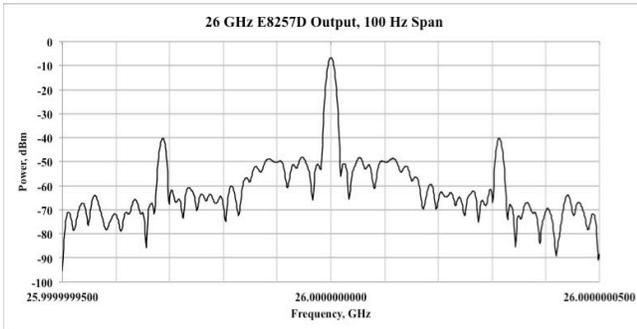
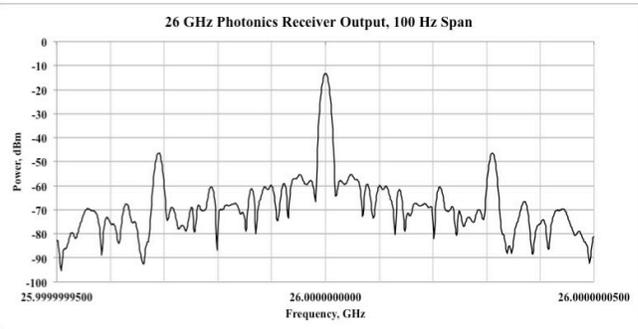

Figure 15. 26.0 GHz output from E8257D synthesizer, 100 Hz span. Note 32 Hz spurs. $P_{out}$ = -6.7 dBm.

Figure 16. 26.0 GHz output from photonics receiver assembly, 100 Hz span. Performance is very similar to synthesizer output. $P_{out}$ = -13.4 dBm.

### 3.3 Secondary LO and Clock Generation

The 10 MHz reference output from the photonics receiver assembly is distributed to several instruments in the receiver cabin, two of which are shown in Figure 17 in relation to the receiver WCA. The backend reference assembly shown in Figure 18 uses the 10 MHz reference to produce and distribute 3.85 GHz and 8.15 GHz LOs for second down conversion, and 2.048 GHz clock for the digitizers. The signal test source reference assembly shown in Figure 19 contains a 31.5 MHz first LO offset generator (FLOOG) that is locked to 10 MHz and distributed to the receiver WCAs along with the 18-31 GHz Ref-1 LO from the photonics receiver. The RF switches within the signal test source assembly are used to route the Ref-1 and Ref-2 FLOOG signals to one of three receiver WCAs. The Ref-1 signal is multiplied-by-4 and mixed with a YIG multiplied-by-6 to produce a 31.5 MHz difference frequency within the WCA. The difference frequency is phase locked to the 31.5 MHz Ref-2 FLOOG. The final LO frequency is defined as follows:

$$F_{1st\ LO} = N*[4 * Ref\text{-}1 + 31.5\ MHz], \qquad (1)$$

where N is the final multiply ratio provided within the CCA and is a value of 1 for the 86 GHz receiver and 3 for the 200 and 345 GHz receivers. Note the large multiply ratio of 12 requires the Ref-1 signal to be of high quality with low phase noise performance.

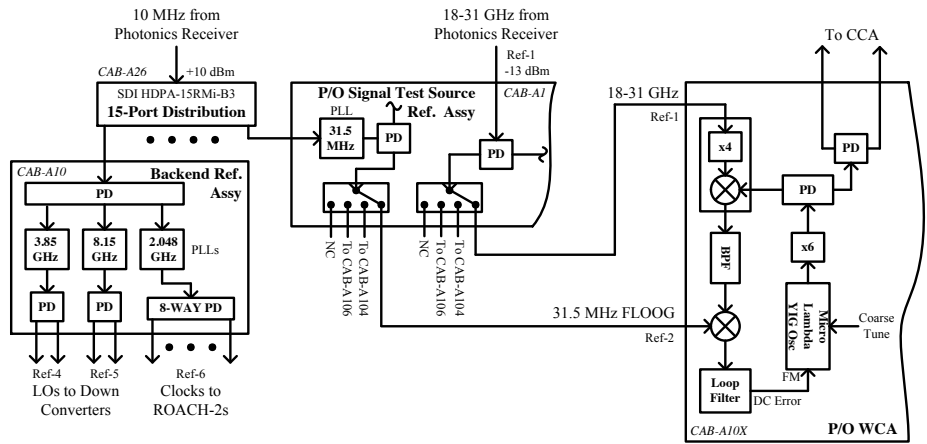

Figure 17. Secondary LO and clock distribution in receiver cabin. The 10 MHz reference from the photonics receiver is distributed to phase lock sources and the 18-31 GHz LO is used to lock the WCA.

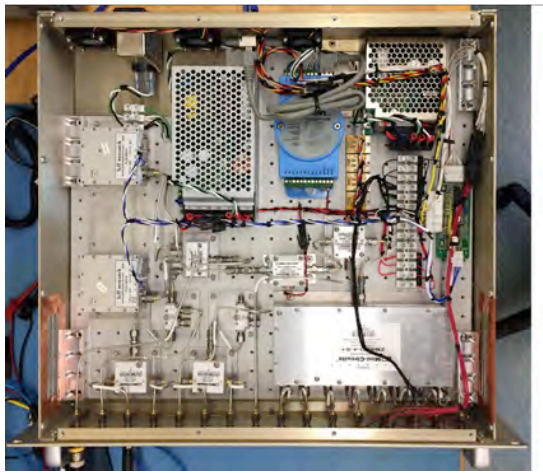

Figure 18. Backend reference assembly. The 3.85 GHz and 2.048 GHz PLS' are shown on the left, 8.15 GHz PLS is missing in the photo.

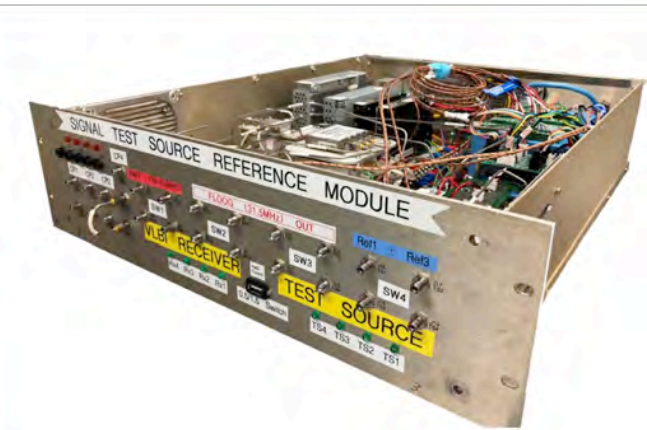

Figure 19. Signal test source reference assembly. Generates 31.5 MHz FLOOG signal and distributes this and the 18-31 GHz LO to the receivers.

## 4. DOWN CONVERSION, DIGITIZATION AND RECORDING

A functional description of the IF processor and baseband distribution assemblies is provided in Figure 20. The GLT system utilizes two IF processors assemblies shown in Figure 21 to accept and process four sets of 4-8 GHz intermediate frequency (IF) signal inputs from the receivers. Each IF processor has the capability to select the IF outputs from one of three receivers. A variable gain amplifier is used in conjunction with a power meter (IF Mon to Switch Matrix) to achieve the desired mixer RF port drive level. The 4-8 GHz signal is power divided (PD) and separately band-pass filtered (BPF) to produce two blocks of 4-6 GHz and 6-8 GHz. These blocks are down converted to baseband using LOs of 3.85 GHz and 8.15 GHz to produce four sets of 0.15-2.15 GHz baseband signals. The signals are low pass filtered (LPF) to meet the Nyquist sampling criteria and level controlled using variable attenuators to satisfy the drive levels into the digitizers. Two sets of baseband distribution plates were added to support slope equalization (EQ) and signal division. All of the controls for the switches, amplifier gains, and variable attenuation are remotely controllable over the network.

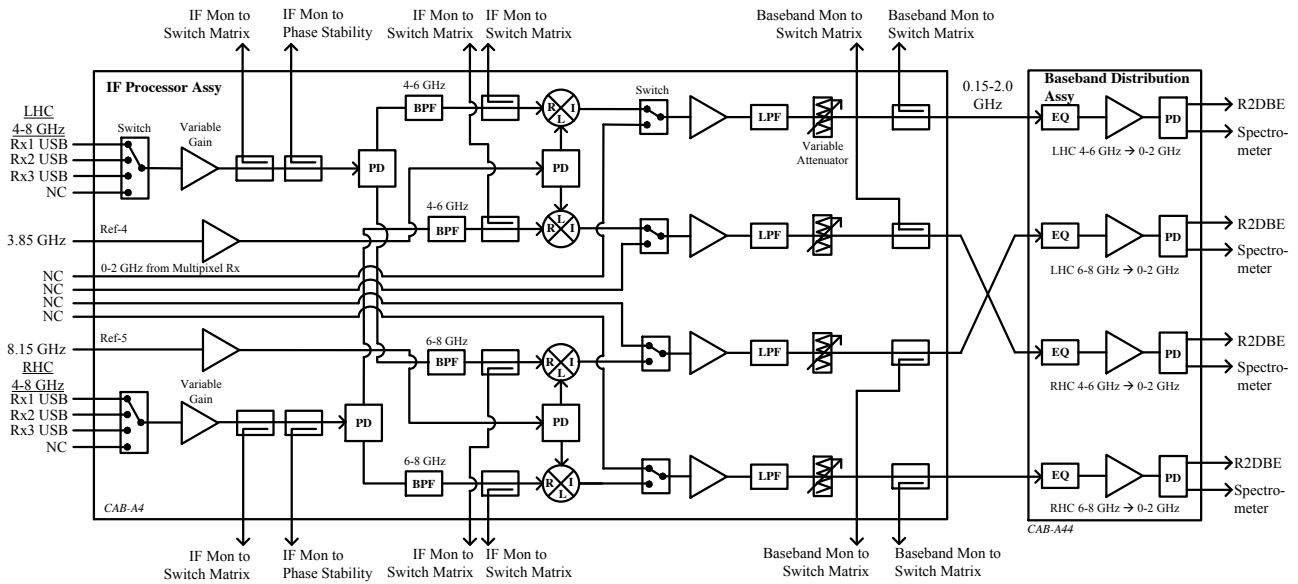

Figure 20. IF processor and baseband distribution assemblies. Accepts 4-8 GHz IFs from the receiver and provides signal leveling, frequency down conversion, and filtering for digitization within the ROACH-2 units. The baseband distribution assembly provides slope equalization and signal division to the R2DBE and spectrometer units.

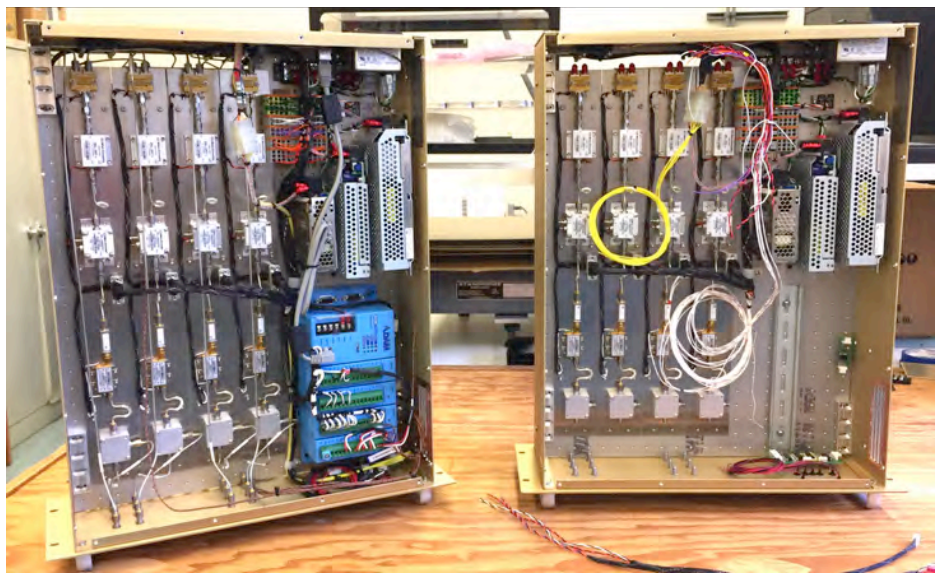

Figure 21. IF processor units under construction. Four IF and baseband paths are visible on the left side of the units. DC power supplies are above the blue monitor and control unit.

A high level system block of the digitization and recording of the baseband signals is shown in Figure 22. The system consists of eight ROACH-2 (reconfigurable open architecture computing hardware, version-2) units, four to support USB and four for LSB (not shown) [5]. Figure 23 shows a photo of ROACH-2 unit that consists of two 5 Gsps 8-bit analog-to-digital converter (ADC) circuit boards [6] attached to the main ROACH-2 board which houses a Xilinx Virtex-6 FPGA. The FPGA permits the unit to be programmed to support an assortment of user-defined functions. The GLT utilizes two separate ROACH-2 configurations that consist of the R2DBE (ROACH-2 digital back-end) and autocorrelation spectrometer. The R2DBE provides the functions of digitization and data formatting for transmission over 10 Gigabit Ethernet (10 GbE) to the Mark-6 data recorders [7] located within the VLBI container. Each ADC has a sampling rate of 4.096 Gsps and the 8-bit quantization is truncated to 2-bits resulting in a transmission data rate of 8.192 Gbps (excludes overhead). The total recording rate into the Mark-6 recorders for both LHC and RHC polarizations and USB and LSB is 64 Gbps. Figure 24 shows a photo of the Mark-6 data recorders in the VLBI trailer.

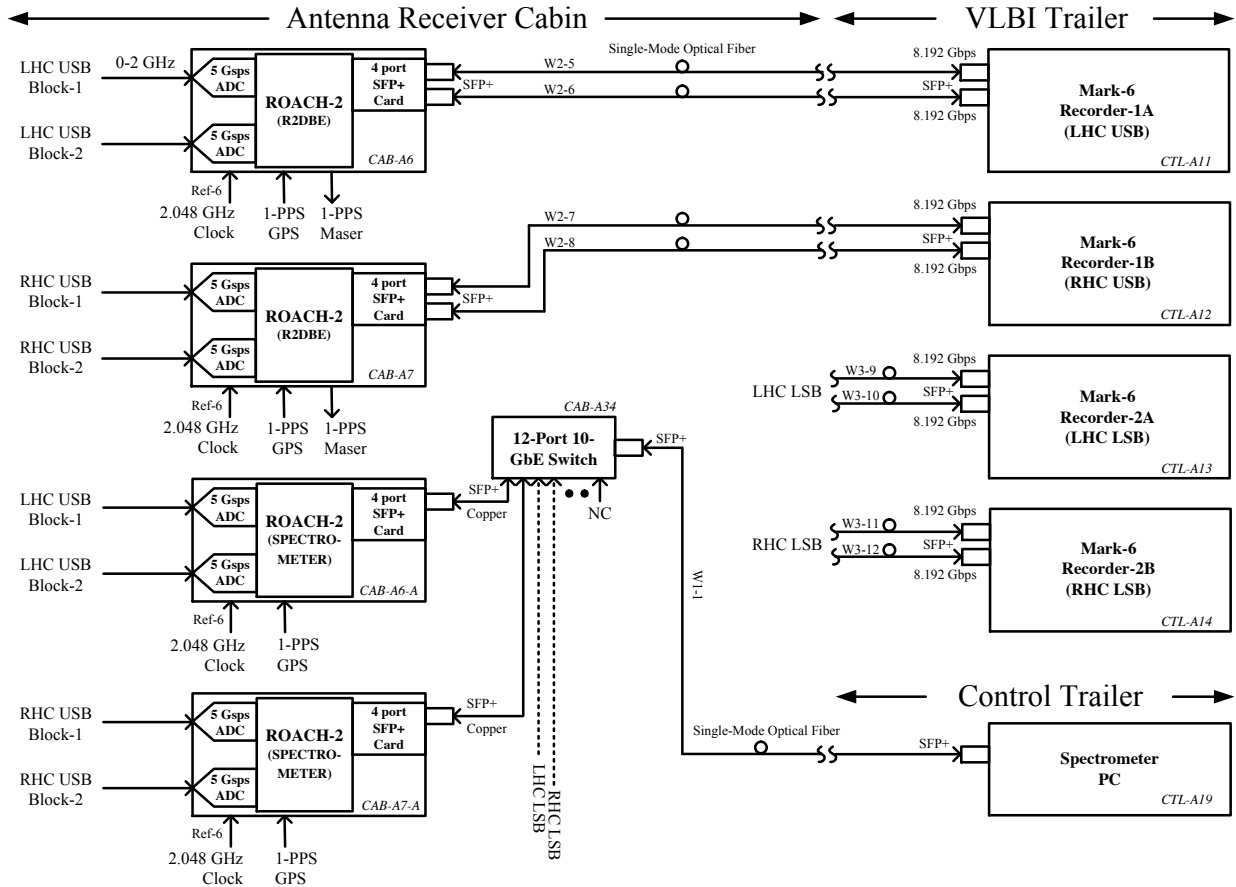

Figure 22. Top level block diagram of GLT digital back end for USB. Two ROACH-2 units are configured as R2DBEs for digitization and transmission to the Mark-6 data recorders located within the VLBI trailer. The lower two ROACH-2 units are configured as autocorrelation spectrometers with the data processed by a spectrometer PC located in the control container. A total of four R2DBEs and four spectrometers are required to support both sidebands. Transmission of the data is over single-mode fiber optic cables.

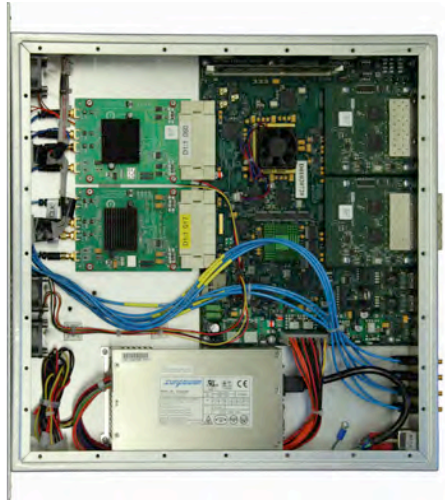
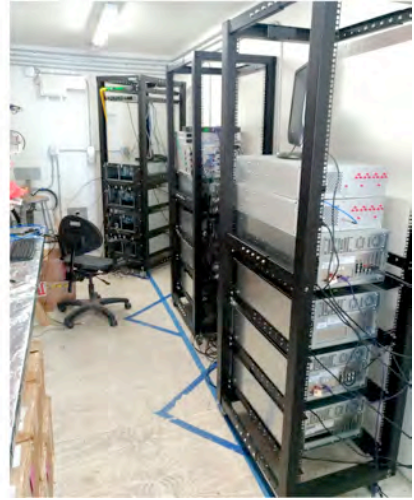

Figure 23.  ROACH-2 chassis.  Two ADC board (light green) are shown attached to ROACH-2 board.  10 GbE network interface cards are on upper right.

Figure 24. VLBI trailer with Mark-6 recorders mounted in racks.

## 5. CALIBRATION AND DIAGNOSTIC TEST SYSTEM

VLBI operations require the recording of large amounts of data by the Mark-6 recorders then sending the hard drives to a central facility for post processing with data from other telescopes.  This process typically requires several months and problems encountered during the actual VLBI observations isn't revealed until much later in time.  With this in mind, and the fact that the GLT is located at a remote site, a number of calibration and diagnostic tests were designed into the system.

### 5.1  10 MHz Maser Performance Monitor

One critical parameter for millimeter VLBI operations is the purity and stability of the 10 MHz reference from our T4 Science iMaser 3000 (serial number 118).  A final LO of 221.0 GHz requires this 10 MHz signal to be scaled by a factor of 22,100 and minute fluctuations of the 10 MHz reference will induce large undesirable effects on the final LO.  The T4 Science unit itself has a web based telemetry window that alerts the user if any of its internal parameters are outside of the specification tolerances.  As an additional check, we use a Symmetricom TSC 5120A phase noise test set to compare the T4 Science 10 MHz output against a separate 10 MHz Oscilloquartz model 8607 oven controlled crystal oscillator that was loaned to the GLT project from MIT Haystack observatory.  The Allan standard deviation performance for our maser was factory tested to be 5e-14 and 8e-15 at 1 and 10 seconds, respectively.  The typical performance for the Oscilloquartz crystal is 0.8e-13 at both 1 and 10 seconds.  The phase noise and Allan standard deviation plots collected on January 19, 2018 are shown in Figures 25 and 26.  The measured Allan standard deviation is 1e-13 at 1 and 10 seconds and is very close to the performance of the Oscilloquartz crystal. It can therefore be indirectly inferred from these measured values that the maser performance must be significantly better than 1e-13.

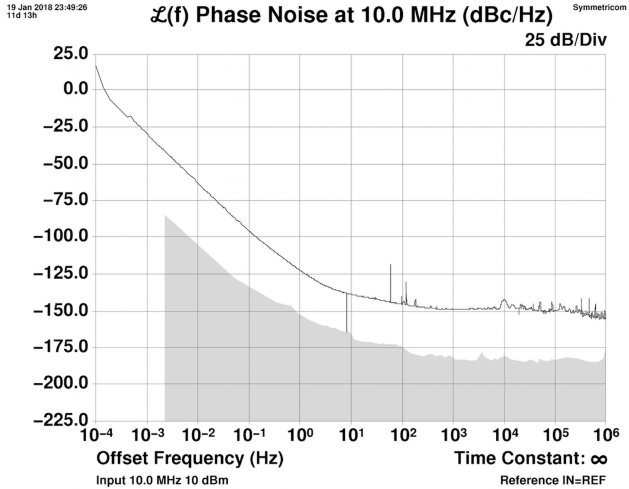

Figure 25. 10 MHz phase noise. Differential phase noise between T4 Science maser and Oscilloquartz OCXO.

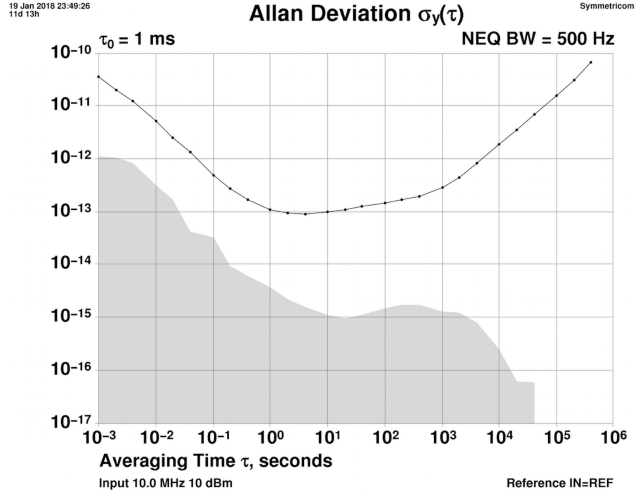

Figure 26. 10 MHz Allan Deviation. Differential Allan deviation between T4 Science maser and Oscilloquartz OCXO.

### 5.2 Round Trip LO Phase Monitor

Referring back to Figure 2, the round trip phase monitor utilizes a 15.75 GHz pilot tone that is summed together with the 18-31 GHz LO and 100 MHz maser reference prior to optical transmission at a wavelength of 1544.5 nm. Within the antenna receiver cabin, the RF pilot tone is separated out from the two other signals using a band-pass filter and is optically re-modulated at wavelength of 1561.4 nm for transmission back to the maser house. The two DWDM add/drop filters permit only the desired optical wavelengths to arrive at their respective Optilab LR-30 receivers. The direct and round trip 15.75 GHz pilot tones are mixed with 15.76 GHz to produce a pair of 10 MHz difference frequencies. A Keysight 8508A vector voltmeter is used to monitor the round trip amplitude and phase stability of the pilot tone. We expect to collect the LO phase stability data when the final LO photonics transmitter assembly is completed and installed in Thule Air Base during September 2018.

### 5.3 Millimeter-Wave Test Tone Generation

The signal test source reference assembly shown in Figure 27 provides the function of distributing the 18-31 GHz Ref-1 and 31.5 MHz Ref-2 FLOOG signals to the three receivers via the pair of SP4T switches. This chassis also generates the LO signals to drive the Pacific Millimeter harmonic mixers shown in the upper left of the figure. Each of the three harmonic mixers are mounted to individual arms that move in and out of their respective receiver beams and is shown in Figure 28. A dual output Valon synthesizer phase locked to 10 MHz is used to produce 500 MHz and 1.5 GHz Ref-3 and serves as the frequency offset for the test tone to the receivers. This Ref-3 signal is mixed with the 18-31 GHz Ref-1 signal using a 90 degree quadrature hybrid and I/Q mixer to produce the desired sum frequency of Ref-1 + Ref-3 and suppresses the undesired difference frequency. The test tone frequency is defined as follows:

$$F_{\text{Test Tone}} = N*4*[*\text{Ref-1} + \text{Ref-3}], \qquad (2)$$

where N and Ref-3 are 1 and 1.5 GHz, respectively for the 86 GHz receiver, and 3 and 500 MHz, respectively for the 230 GHz and 345 GHz receivers. As an example for receiver-1 using (1), a first LO of 221.100000 GHz requires Ref-1 to be 18.417125 GHz. The resultant test tone will be 227.005500 GHz producing a receiver IF tone output of 5.905500 GHz. As a consequence of the design, the IF test tone output for the 230 GHz and 345 GHz receivers is 5.905500 GHz, and for the 86 GHz receiver is 5.968500 GHz.

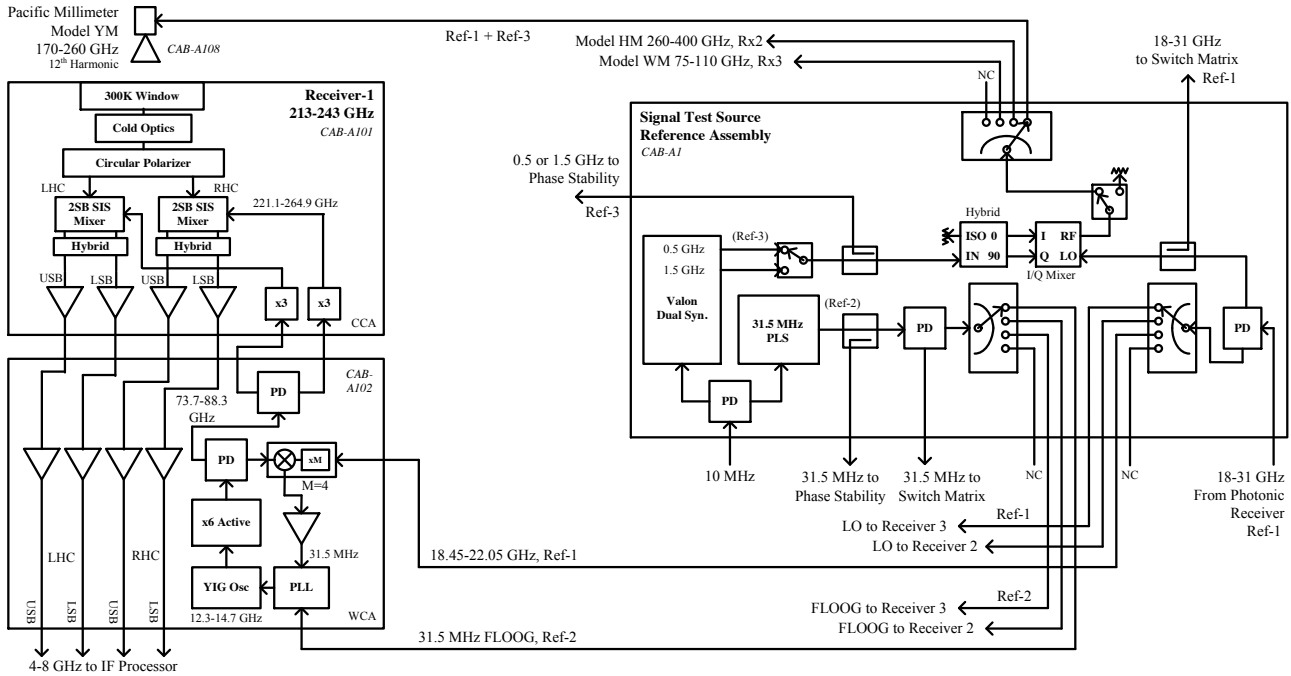

Figure 27. Millimeter-wave test tone generation. The signal test source unit distributes the 18-31 GHz Ref-1 and 31.5 MHz Ref-2 FLOOG to the three receivers. This unit also generates the offset LO to the harmonic mixers for the millimeter tone injection into the receivers.

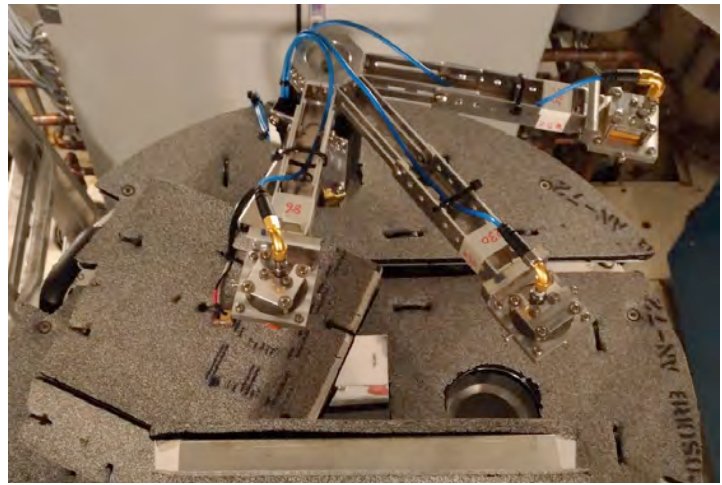

Figure 28. Millimeter wave test tone arms. Left to right are ambient load, 86 GHz, 230 GHz, and 345 GHz test tone arms fixed to a rotary stage.

## 5.4 Test Tone Phase Stability Monitor

A functional block diagram and photo of the continuum detector and phase stability monitor assembly are shown in Figures 29 and 30. The 4-8 GHz signal from the IF processor assembly is band-pass filtered and mixed with a locally derived 5.950 GHz Ref-7 LO that is phase locked to the 10 MHz reference. When the receiver is looking at the millimeter wave test tone source, the resultant tone from the mixer output is 44.5 MHz for the 230 GHz and 345 GHz receivers, and 18.5 MHz for the 86 GHz receiver. The circuit network in the dashed upper right box mixes the 44.5

MHz tone with 94.5 MHz (3 x Ref-2) to produce a difference frequency of 50 MHz. Similarly, the 18.5 MHz tone is mixed with 31.5 MHz (Ref-2) to produce a sum frequency of 50 MHz. The 50 MHz test tone output from the mixer is band-pass filtered and sent to the input port of the vector voltmeter. The 500 MHz / 1.5 GHz Ref-3 signal generated within the signal test source reference assembly is multiplied by 12 or 4 to produce a desired IF of 6.000 GHz and is subsequently mixed with the 5.950 GHz Ref-7 to produce the 50 MHz tone to the reference port of vector voltmeter. A diagram illustrating the phase cancellation of the various reference LOs is provided in Figure 31 and shows that the resultant phase seen by the vector voltmeter represents the stability of the receiver. The graph presented in Figure 32 represents the long-term phase (B-A) and amplitude (B/A) stability of the 230 GHz receiver over an eight hour time period that was started at 11:30pm April 21, 2018. The phase gradually increased by approximately 50 degrees over a period of four hours (0.21 degrees/minute) then flattened out for the remaining four hours. A 50-degree phase movement at a test tone frequency of 227.0055 GHz corresponds to a physical movement of 0.184-mm between the tone source located on the swing arm above the receiver and the receiver itself (refer to Figure 28). This stability result is sufficiently good for VLBI applications.

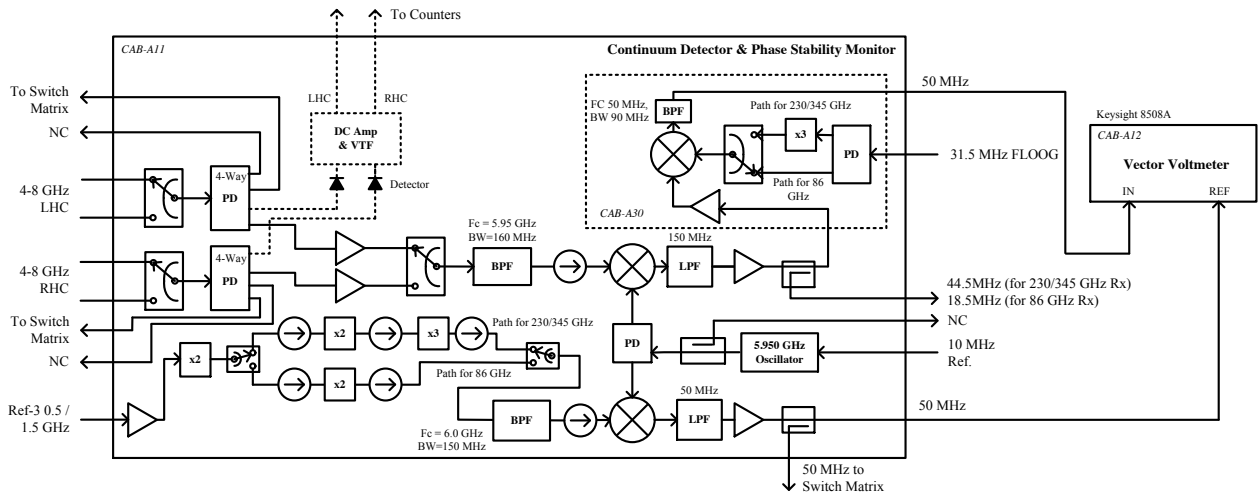

Figure 29. Continuum detector and phase stability monitor assembly. One of four 4-8 GHz IF signals containing the desired test tone is selected via switches and down converted to 44.5/18.5 MHz using a 5.95 GHz LO. This signal is mixed with 94.5/31.5 MHz (upper right dashed box) to produce 50 MHz test tone to the vector voltmeter. A 50 MHz reference tone for the vector voltmeter is generated by mixing (12 x 0.5)/(4 x 1.5) GHz with the 5.95 GHz LO. The two detectors from the LHC and RHC polarization inputs are DC amplified and voltage-to-frequency converted for synchronous continuum detection.

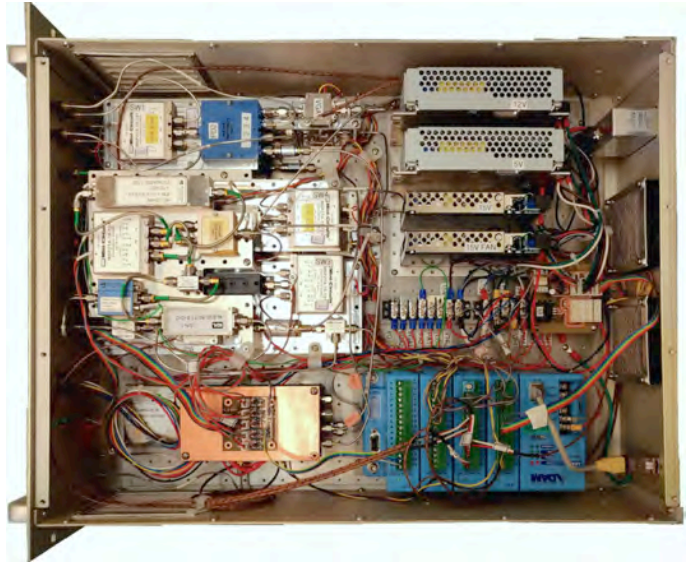

Figure 30. Continuum detector and phase stability monitor assembly. This assembly is rack mounted in the antenna receiver cabin.

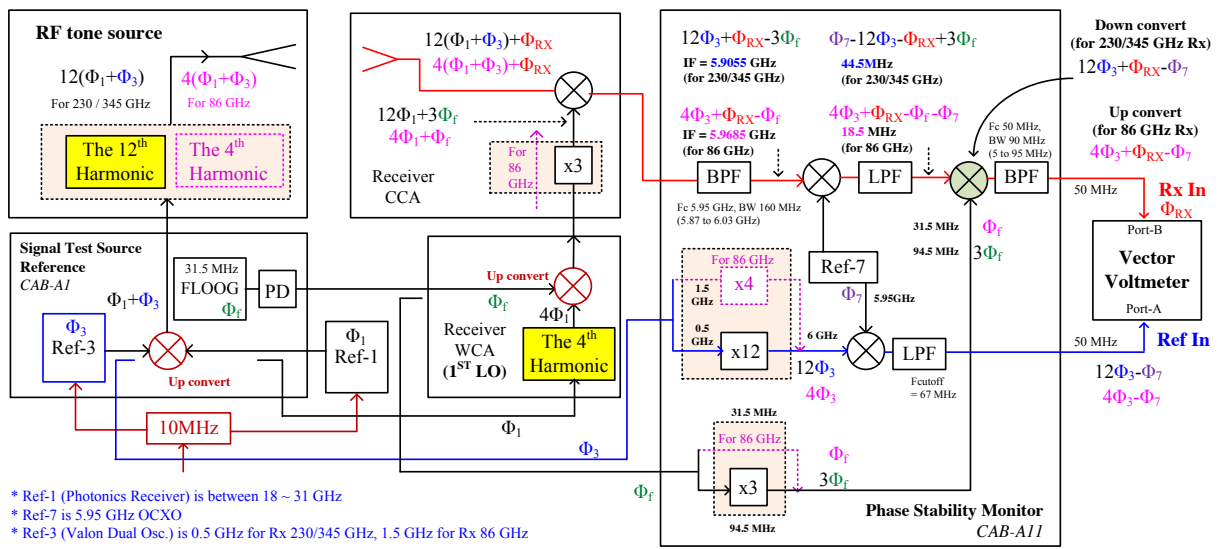

Figure 31. Phase cancellation diagram. Phase variations in Ref-1 (18-31 GHz), Ref-2 (31.5 MHz FLOOG), Ref-3 (0.5/1.5 GHz), and Ref-7 (5.95 GHz) are cancelled and leave only the phase variations caused by the receiver system.

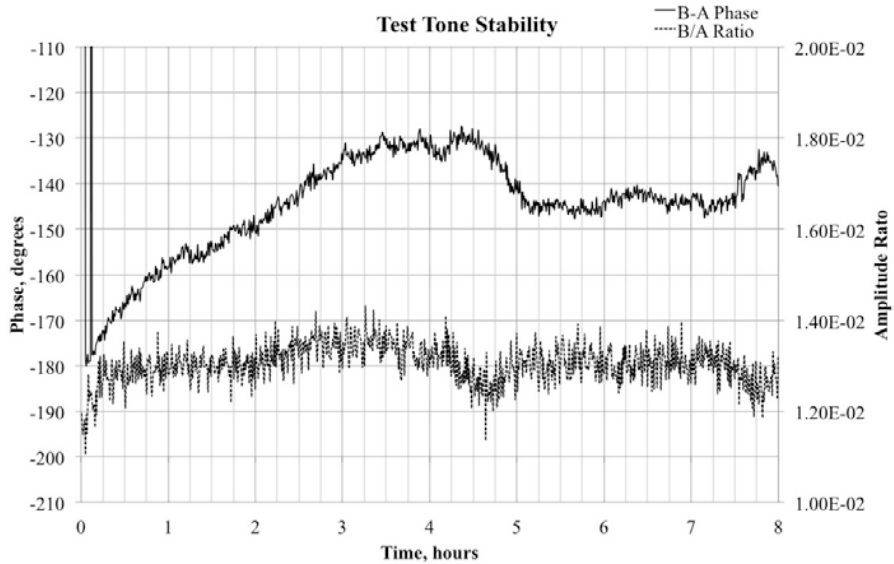

Figure 32. 230 receiver test tone phase and amplitude response. A 227.0055 GHz test tone was generated and transmitted to the receiver where it was mixed with a first LO of 221.1000 GHz to produce an IF of 5.9055 GHz. This IF tone was mixed and converted to 50 MHz within the continuum detector & phase stability monitor unit then phase and amplitude compared to a 50 MHz reference.

### 5.5 Switch Matrix Assembly

The GLT receiver cabin instrumentation system utilizes eight Keysight E4413A power sensors (0.05-26.6 GHz) connected to four N1914A power meters for monitoring signal levels. This system also employs a Keysight N9010A spectrum analyzer (9 kHz – 44 GHz) for monitoring broadband IF spectra and narrowband LO signals. Monitor tests points have been built into each of the custom assemblies and at present we have 31 individual signals to monitor with either the power meter and/or the spectrum analyzer. In order to accomplish these monitoring scenarios remotely, two RF switch matrix assemblies have been designed to route the various monitor test points to either one of eight power meters or the spectrum analyzer. Figure 33 represents the interconnect of the two RF switch matrix chassis in relation to the power meters and spectrum analyzer. The units are currently under construction, Figure 34, and will be installed into the receiver cabin rack during the summer of 2018.

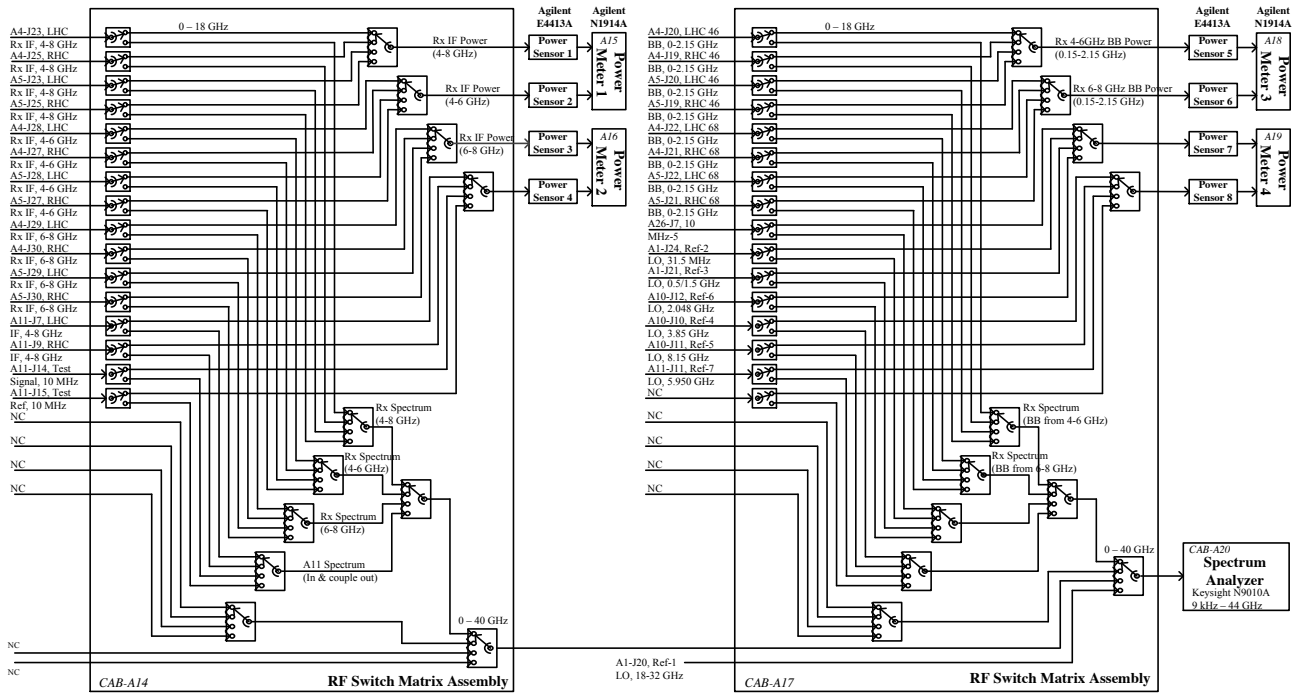

Figure 33. RF switch matrix interconnection diagram. Each unit contains sixteen SPDT and eleven SP4T RF switches for remote monitoring of up to 42 signals within the receiver cabin.

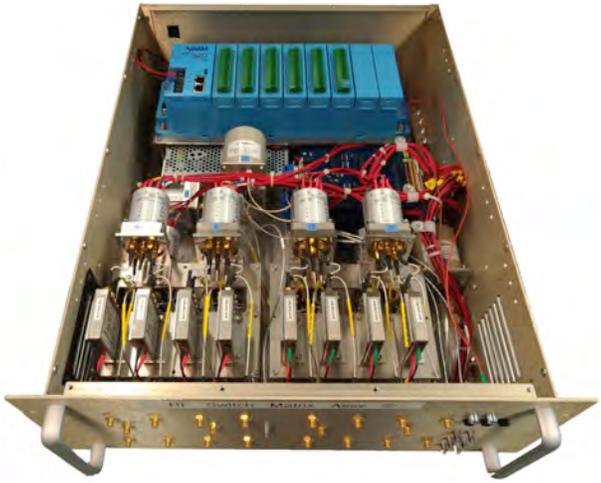

Figure 34. RF switch matrix assembly. Two units are currently under construction and will be deployed into the receiver cabin toward the end of summer 2018.

## 6. FIBER OPTICS SYSTEM

All RF reference signals and data communications to and from the telescope are transported over low temperature single-mode AFL fiber optic cables capable of supporting 1310-nm and 1550-nm optical windows. This cable is designed to withstand operation down to -65 C and consists of a 1.32-mm OD 316 stainless steel tube that contains four Verrillon VHS100 polyimide coated fibers. The stainless steel tube is spiral wrapped by #12 316 stainless steel wires and a 3.80-mm OD HDPE outer jacket. We selected this small diameter 4-fiber cable because of the small dynamic bend radius of 132-mm (5.2 inches). The GLT system consists of twelve fiber optic cables routed from the Control trailer, VLBI trailer, and Maser house to three separate locations within the telescope. Figure 35 illustrates the routing of the cables between the trailers and telescope. Note that all twelve of the fiber cables must pass through the telescope azimuth wrap with eight also passing through the elevation wrap. The current system utilizes seven of the twelve cables with each active cable fiber terminated with SC/APC connectors as shown in Figure 36. The remaining five spare cables will be terminated with SC/APC connectors during the summer of 2018. Figure 37 shows the terminal equipment at each end of the active fibers.

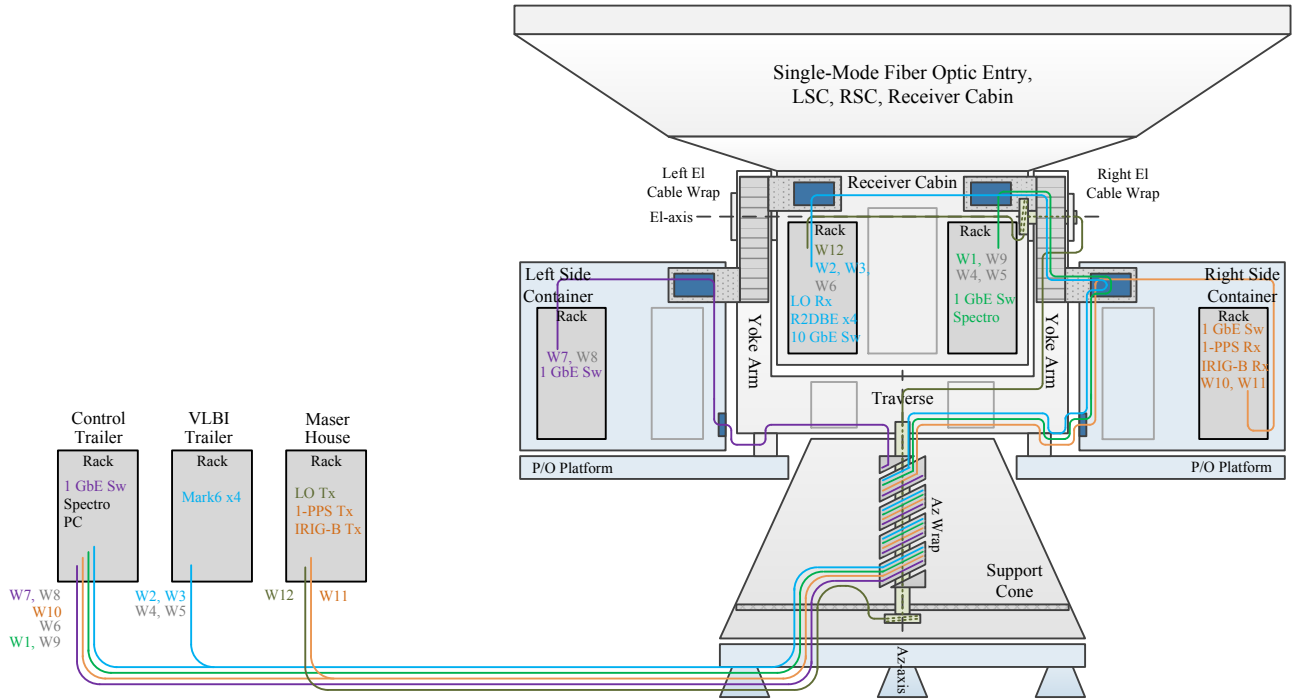

Figure 35. Routing of fiber optic cables between trailers and telescope. Each of the cables contains 4 single mode fibers that are terminated in racks at both ends. All cables must traverse through the azimuth wrap of the telescope.

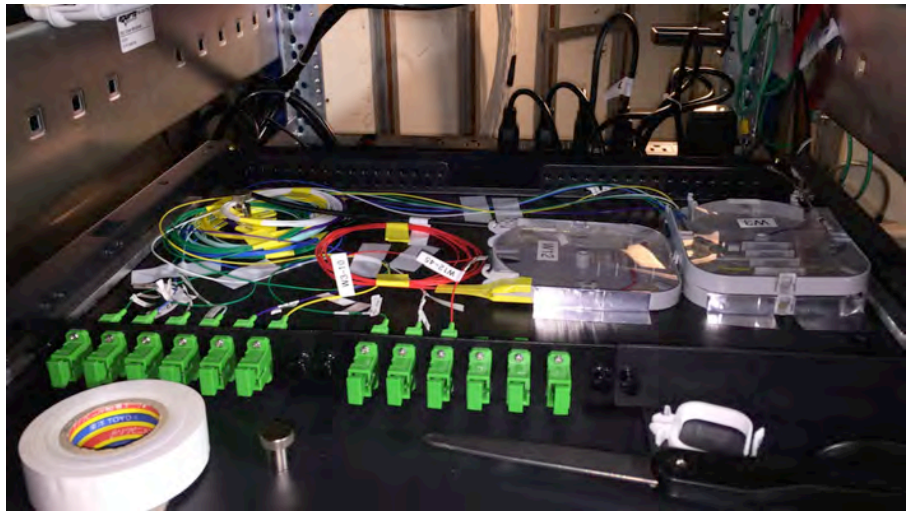

Figure 36. Fiber termination chassis located in terminal racks. This photo is of the receiver cabin left rack, which provides terminations for cables W2, W3, and W12.

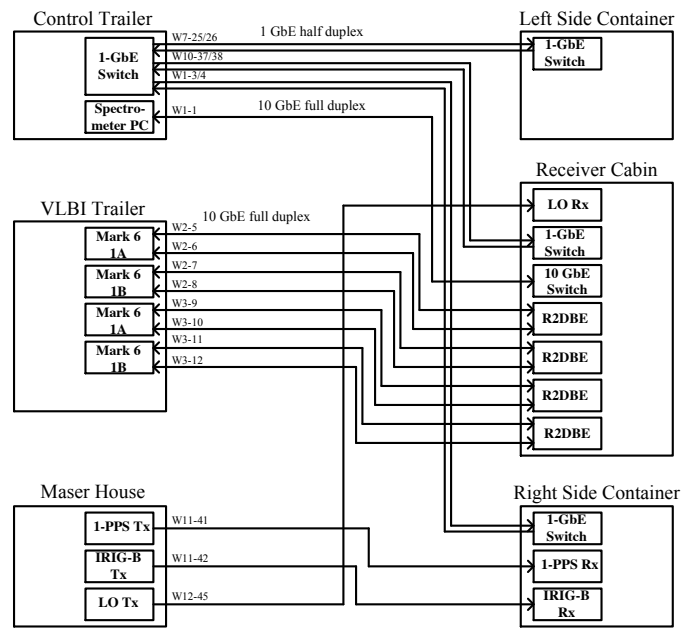

Figure 37. Terminal equipment at the end of the active fibers. The 1 GbE network utilizes half duplex communication and requires two fibers, one each for transmit and receive. The 10 GbE network utilizes full duplex bidirectional communication and requires a single fiber for both transmit and receive.

## CONCLUSION

The initial integration and test of the electronics instrumentation system was performed at the ASIAA facility in Hilo, Hawaii, during the summer 2017. Working with our partners at the East Asian Observatory, the GLT instrumentation system was transported and integrated into the JCMT on Maunakea. Astronomical first light was obtained during September 2017. Shortly after, connected interferometric fringes were obtained between the JCMT and SMA telescope facilities using this system and validated the design concepts for several of the new GLT instruments and assembly units. The next phase involved transporting the hardware to the GLT site in Thule Air Base, Greenland. Installation and test began in early November 2017 and first light was obtained at the end of December. The science team has continued with telescope commissioning work at Thule Air Base and has recently participated in global VLBI observations at 86 GHz and 230 GHz. Our plans for 2018 are to complete and install the finalized instruments and assemblies and to refine the automated processes to support remote operations.

## ACKNOWLEDGEMENTS

The ASIAA's funding for the GLT is partially supported by the Ministry of Science and Technology, MOST funding codes: 99-2119-M-001-002-MY4, 103-2119-M-001-010-MY2, and 106-2119-M-001-013, for Taiwan's participation in the ALMA-NA project. The realization of the GLT provides a unique baseline for the ALMA's VLBI capability. Support for the Hydrogen Maser frequency standard used for VLBI at the Greenland Telescope was provided through an award from the Gordon and Betty Moore Foundation (GBMF-5278). The GLT project thanks the strong support from the National Chun-Shan Institute of Science and Technology. We also thank the National Astronomical Observatory of Japan for their support in the receiver instrumentation, the US National Science Foundation, Office of Polar Programs for effective support in logistics, and to the United States Air Force, 821[st] Air Base Group, Thule Air Base, Greenland for use of the site and access to the Base and logistics chain. The project appreciates the development work done by Vertex Antennentechnik GmbH and ADS international. We'd also like to thank Atunas, an outdoor wear company in Taiwan for the arctic clothing. Lastly, we thank the National Radio Astronomy Observatory and MIT Haystack Observatory for their support in the acquisition of the ALMA-NA prototype in order to repurpose it for Greenland deployment.